\documentclass[12pt]{article}
\usepackage{ssi}
\usepackage{times}
\usepackage{pennames}
\usepackage{epsf}

\def\ipb{pb$^{-1}$}
\def\etal{{\it et al.}}
\def\Mz{M_{\mathrm{Z}}}
\def\Gz{\Gamma_{\mathrm{Z}}}
\def\Gzinv{\Gamma_{\mathrm{invis}}}
\def\Nnu{N_{\nu}}
\def\Gznusm{\Gamma_{\nu\nu}(\mathrm{SM})}
\def\Mw{M_{\mathrm{W}}}
\def\Gw{\Gamma_{\mathrm{W}}}
\def\ffbar{f\overline{f}}
\def\qqbar{\mathrm{q}\overline{\mathrm{q}}}
\def\bbbar{\mathrm{b}\overline{\mathrm{b}}}

\def\Apar{{\mathcal{A}}}
\def\Apare{{\mathcal{A}_{e}}}

\def\Aparl{{\mathcal{A}_{\ell}}}
\def\Aparf{{\mathcal{A}_{f}}}
\def\Aparb{{\mathcal{A}_{b}}}
\def\Aparc{{\mathcal{A}_{c}}}
\def\gvf{g_{Vf}}
\def\gaf{g_{Af}}
\def\afb{A_{FB}}

\def\afbb{A_{FB}^b}
\def\afbc{A_{FB}^c}
\def\afbzf{A_{FB}^{0,f}}

\def\alr{A_{LR}}
\def\alrz{A_{LR}^0}

\def\Mh{M_{\mathrm{H}}}
\def\Mtop{M_{\mathrm{t}}}
\def\qqqq{\mathrm{q}\mathrm{q}\mathrm{q}\mathrm{q}}
\def\qqln{\mathrm{q}\mathrm{q}\ell\nu_{\ell}}

\def\sstw{\sin^2\theta_{\mathrm{W}}}
\def\ssteff{\sin^2\theta_{\mathrm{eff}}}
\def\sstefff{\sin^2\theta_{\mathrm{eff}}^f}
\def\ssteffl{\sin^2\theta_{\mathrm{eff}}^{\mathrm{lept}}}

\begin{document}

\title{LEP, SLC and the Standard Model}

\author{D.G. Charlton\thanks{Royal Society University Research Fellow
(to 30 September 2002)
\vskip 0.5in 
\noindent
\copyright\ 2002 DG Charlton}\\ 
School of Physics \& Astronomy \\
The University of Birmingham, Birmingham, B15 2TT, U.K. \\[0.4cm]
}

\maketitle
\begin{abstract}%
\baselineskip 16pt 
The period 1989-2000 provided a huge yield of precise electroweak data
from the LEP and SLC experiments. 
Many analyses of these data are now complete, but others, particularly
of the full LEP-2 data samples, continue.
The main electroweak physics results from these data are summarised,
and stringent tests of the Standard Model are made with the combined
samples.
The direct search for the missing link of the Standard Model, the
Higgs boson, is also briefly reviewed.
\end{abstract}

\section{Introduction: Data Samples}

The Stanford Linear Collider, SLC, was the first electron-positron
collider operating at centre-of-mass energies at and around the \PZz\
pole. 
From the startup in 1989 to the final data collected in 1998, around
20\,\ipb\ of integrated luminosity was accumulated, in very large part
by the SLD detector.
Although the data sample is modest when compared to that from LEP, the
power to probe the Standard Model is greatly enhanced by the
substantial electron beam polarisation, typically 75\% in the later
years of SLC operation.
This unique feature of the \PZz\ data collected by SLD results in the
most precise single measurement of the weak mixing angle, as described
below.

\begin{figure}[htb]
\begin{center}
\mbox{\epsfysize=7cm\epsfbox{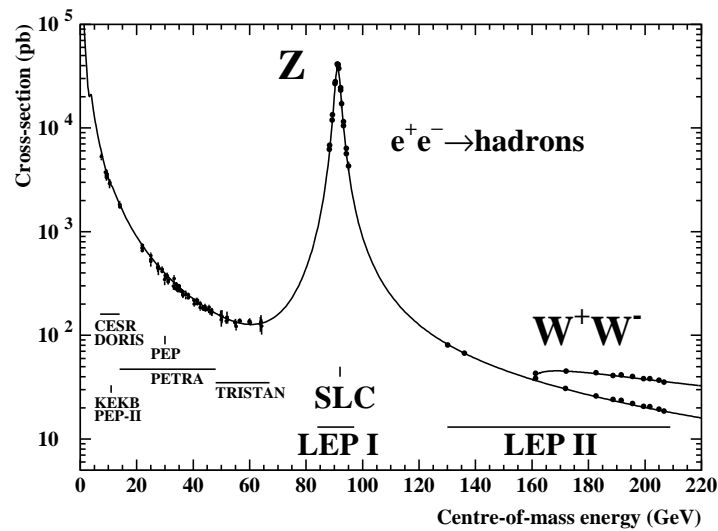}}
\end{center}
\caption{Hadronic \Pep\Pem\ annihilation cross-section from the B
  factories to LEP-2.}
\label{fig:shad_wide}
\end{figure}

The large electron-positron collider, LEP, sited at CERN, also started
taking data in 1989, and each of the four experiments collected
approximately 160\,\ipb\ of data at and around the \PZz\ peak in the
years up to 1995, corresponding to a total of more than 15 million
observed \PZz\ decays. A several year programme to increase the
accelerating voltage resulted in data-taking above the threshold for
\PWp\PWm\ production from 1996 to 2000. At these energies
substantially increased luminosities were also possible, resulting in
approximately 700\,\ipb\ of data being collected in this ``LEP-2'' phase
of operation.
As illustrated in figure~\ref{fig:shad_wide}, the cross-section for 
W-pair production is three orders of magnitudes lower than that at the Z
peak, so that W-pair events collected are numbered in thousands rather
than millions. 

\section{${\mathbf{Z^0}}$ Production and Decays}

\subsection{The ${\mathbf{Z^0}}$  Lineshape}

\begin{figure}[htb]
\begin{center}
\mbox{\epsfysize=7cm\epsfbox{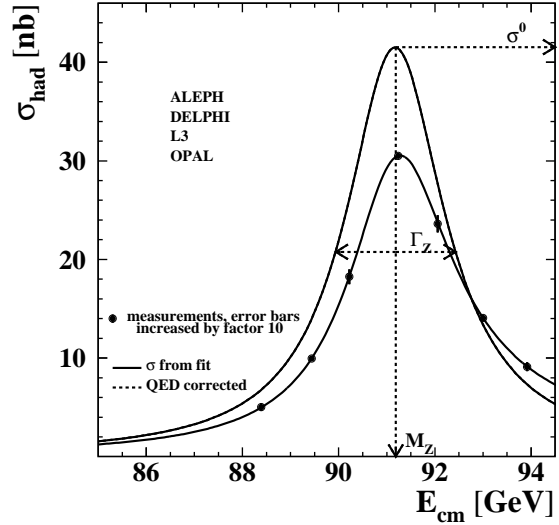}}
\end{center}
\caption{Schematic illustration of the extraction of the \PZz\
  lineshape parameters from the measured cross-sections at
  LEP\cite{leplineshape}.}
\label{fig:zls}
\end{figure}

The measurement of the cross-section for fermion pair production at
and around the \PZz\ pole was used to measure the Z mass and width, as
illustrated schematically in~figure~\ref{fig:zls}. Note that a
substantial correction is needed for the effects of QED initial-state
photon radiation (``ISR'').
The analyses from the four LEP experiments have now been final for
some time\cite{lineshape}, and combining them\cite{leplineshape}
yields:
\begin{eqnarray}
\Mz & = & 91.1875\pm0.0021~\mathrm{GeV} \\
\Gz & = & 2.4952\pm0.0023~\mathrm{GeV} \mbox{\ \ \ \ } .
\end{eqnarray}

\begin{figure}[htb]
\begin{center}
\mbox{\epsfysize=6.5cm\epsfbox{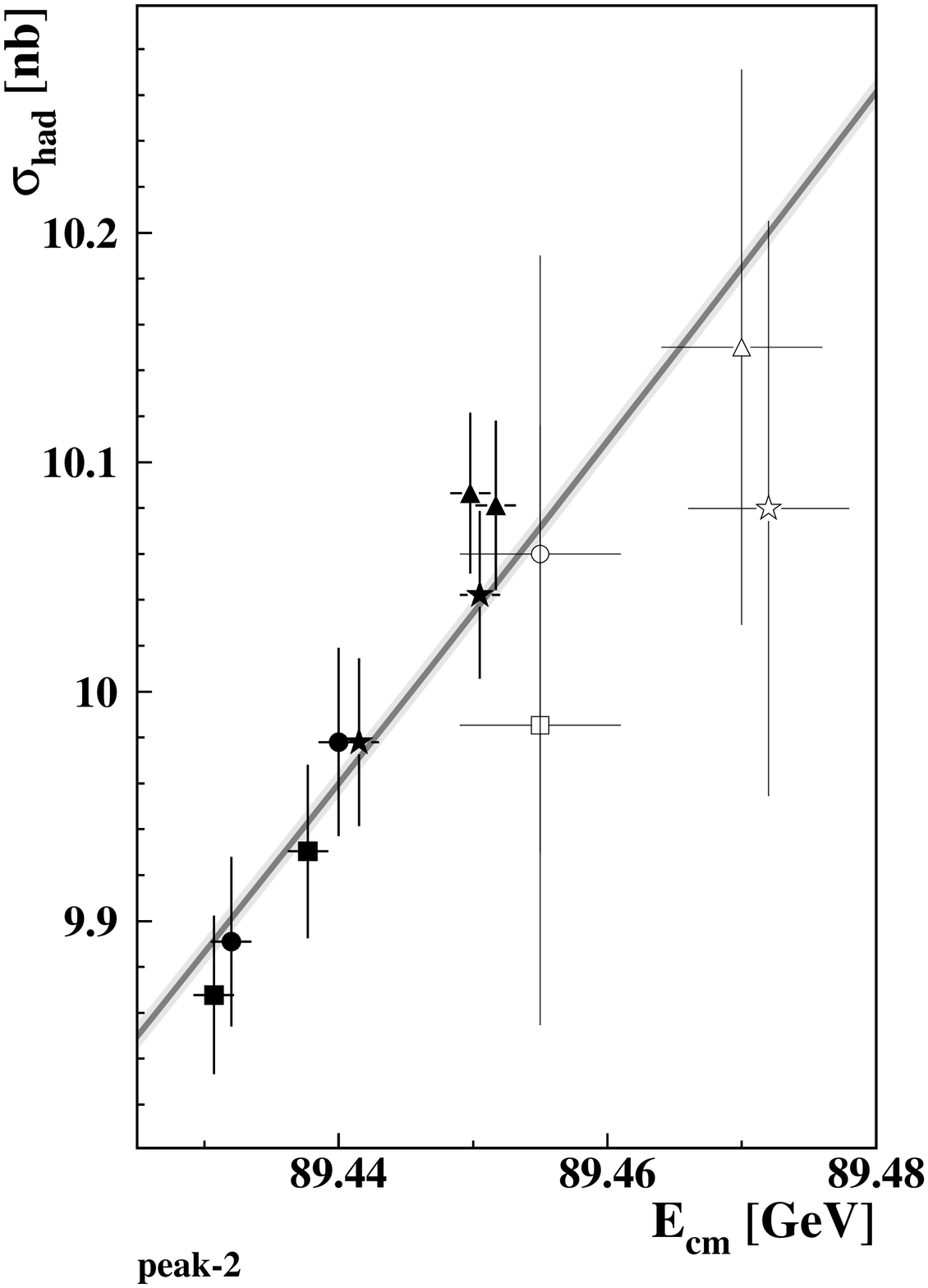}
\epsfysize=6.5cm\epsfbox{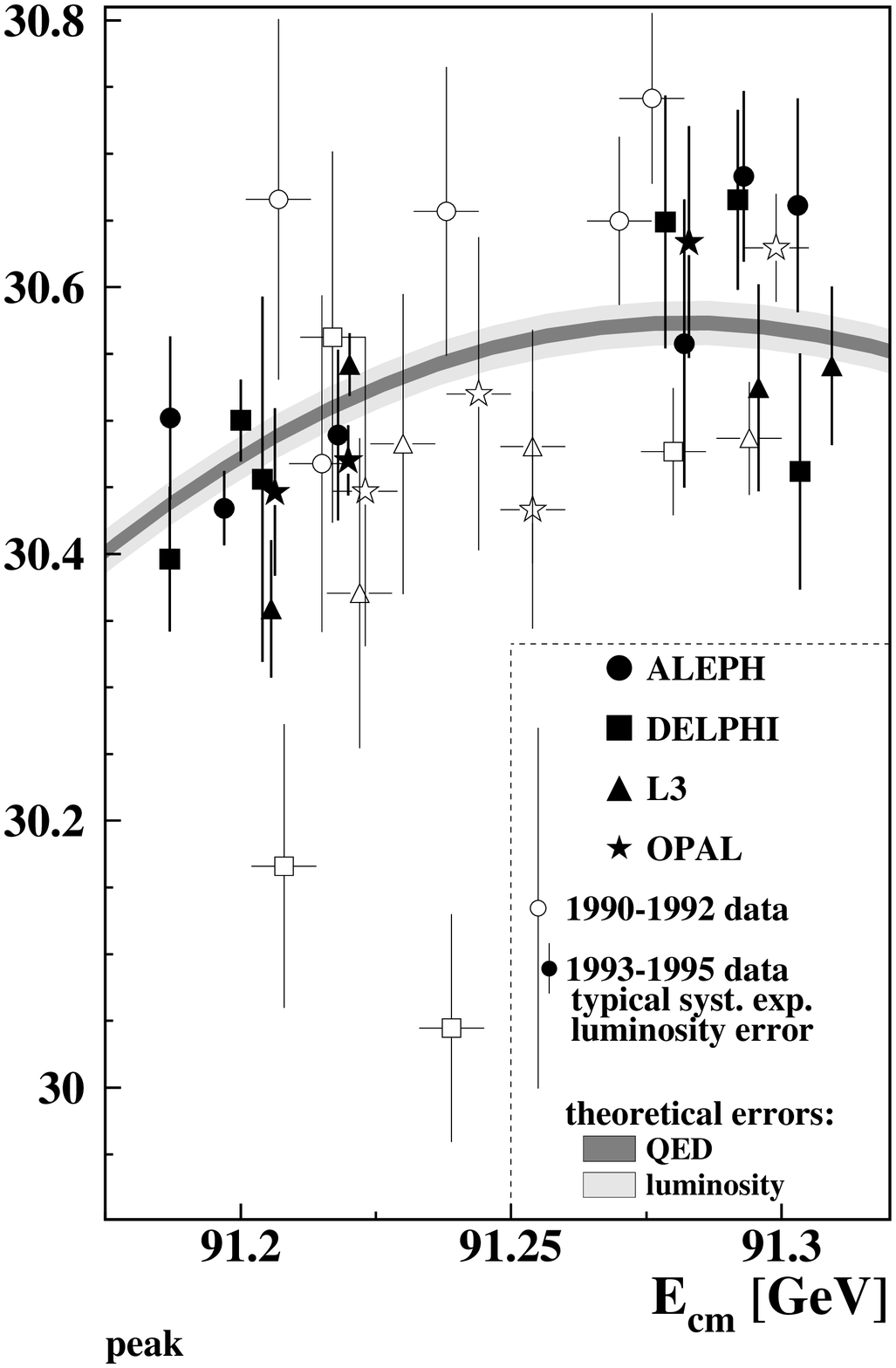}
\epsfysize=6.5cm\epsfbox{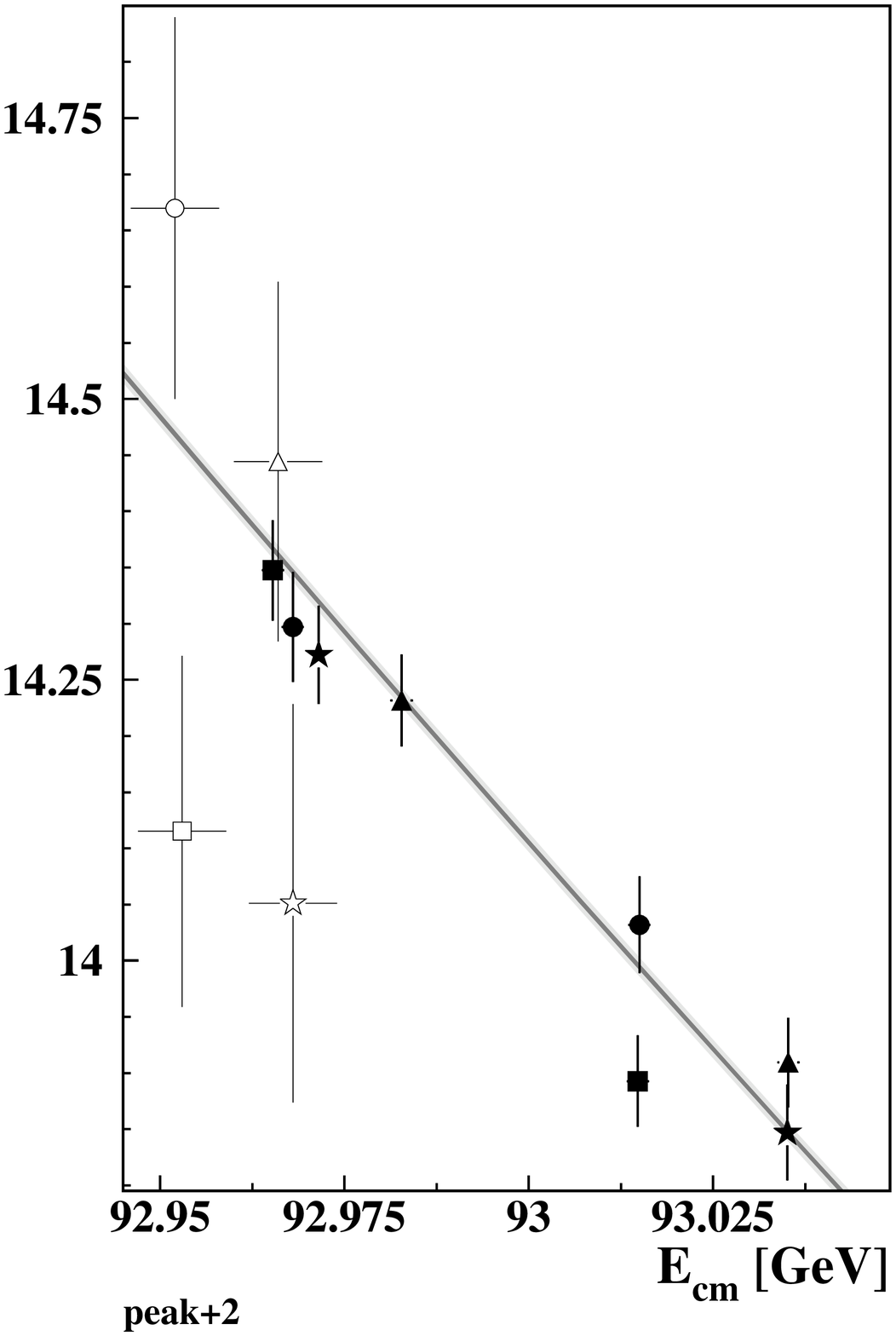}}
\end{center}
\caption{Detail of the hadronic cross-section measurements from LEP
  around the highest statistics points\cite{leplineshape}.}
\label{fig:zhistats}
\end{figure}

The measurement of the cross-sections to different particle types,
either inclusive hadrons, different charged lepton species, or tagged
primary quark flavours, gives access to the individual \PZz\ partial
decay widths, $\Gamma_{\ffbar}$,  via
\begin{equation}
\sigma^0(\PZz\to\ffbar)=\frac{12\pi}{\Mz^2}\frac{\Gamma_{\mathrm{ee}}\Gamma_{\ffbar}}{\Gz^2}
\end{equation}
after correction for ISR effects and for the effects of the t-channel
diagrams in the case of the Bhabha scattering process
$\Pep\Pem\to\Pep\Pem$. 
The extremely high statistics available at the three most precisely
measured points, at the \PZz\ peak and $\pm$2~GeV away, are
illustrated in figure~\ref{fig:zhistats}.

Further use is made of the Z lineshape results to derive the
partial decay width
of the \PZz\ to unobserved (invisible) decays, $\Gzinv$:
\begin{equation}
\Gz = (\sum_{\mathrm{visible}~f} \Gamma_{\ffbar}) + \Gzinv
\end{equation}
where the sum over visible decay products uses the hadronic decay
width and the individual charged leptonic ones.
If neutrinos are further assumed to have Standard Model couplings, the
number of light neutrino species may also be determined from
$\Nnu=\Gzinv/\Gznusm$, giving:
\begin{equation}
\Nnu=2.9841\pm0.0083 \mbox{\ \ \ \ } .
\end{equation}

\begin{figure}[htb]
\begin{center}
\mbox{\epsfysize=7.5cm\epsfbox{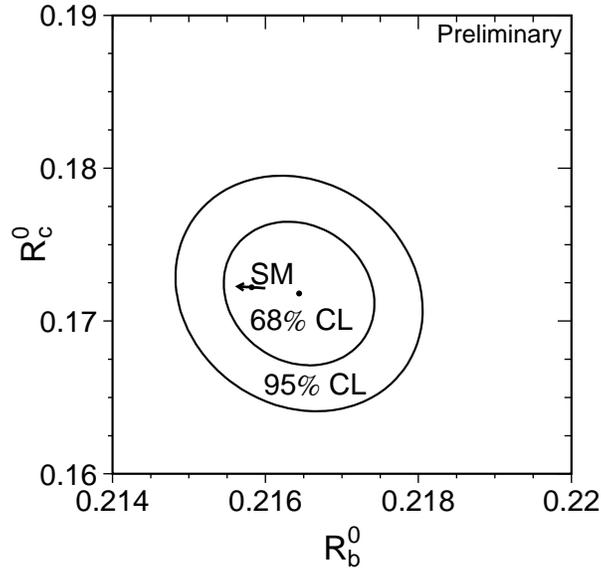}}
\end{center}
\caption{Measured values of $R_b$ and $R_c$ compared with the Standard
  Model expectation (the arrow shows the effect of varying the top
  mass over the range 174.3$\pm$5.1~GeV. 
  Results are corrected to the Z pole (denoted
  by the 0 superscript).}
\label{fig:rbrc}
\end{figure}

Measuring individual Z fermionic decay widths gives access to the
sum of the squares of the vector and axial-vector couplings of
the fermion species to the \PZz,
$\Gamma_{\ffbar}\propto(\gvf^2+\gaf^2)$.
While this is straightforward for the charged lepton flavours, for
quark decay modes it requires high quality flavour tagging: this is
only available in practice for b and c quarks, where combinations of
tags, utilising such properties as the high b mass and lifetime, are
employed. 
The results obtained are usually expressed in terms of the ratios
$R_Q$, defined by
$R_{Q}\equiv\frac{\Gamma_{\mathrm{Q}\overline{\mathrm{Q}}}}{\Gamma_{\mathrm{had}}}$.
Although the performance of such tags can be excellent in terms of
efficiency and purity, especially for b quarks, the precise
performance cannot be simulated adequately. This is circumvented by
the use of ``double tag'' techniques, where the tag is applied
independently to the two hemispheres of an event, allowing the
tagging efficiency and $R_Q$ both to be determined.
Combining the LEP and SLD results\cite{rbrc}, the results for 
$R_b$ and $R_c$ as
shown in figure~\ref{fig:rbrc} are obtained. The Standard Model prediction
describes the data well, in contrast to the situation seven years
ago\cite{rbcrisis},
when the first precise measurements of $R_b$ were available.

\subsection{Asymmetries}

More information about electroweak parameters is contained in
\PZz\ events, basically independent of the cross-section
measurements, via the measurement of asymmetries.
Many asymmetries have been measured at LEP, and a further powerful
probe is possible with the polarised electron beam at SLD.

In the framework of the Standard Model, the various measured
asymmetries -- after correction for ISR effects -- may be written in
terms of the asymmetry parameters $\Apar$ defined by 
$\Aparf \equiv \frac{2\gvf\gaf}{(\gvf^2+\gaf^2)}$, which is simply a
function of the ratio
$\frac{\gvf}{\gaf}=1-4|Q_f|{\sstefff}$.
The effective weak mixing angle $\sstefff$ differs slightly for
different
fermion species, although in practice most asymmetries probe the value
for charged leptons, $\ssteffl$.
For example, the forward-backward asymmetry, defined as the ratio
$\afb=\frac{(\sigma_F-\sigma_B)}{(\sigma_F+\sigma_B)}$, where
$\sigma_{F(B)}$ is the cross-section for fermions $f$ scattering in
the same
(opposite) hemisphere as the original electron beam direction, may be
expressed as $\afbzf=\frac{3}{4}{\Apare\Aparf}$. As before, the zero
superscript denotes correction for ISR effects to Z pole quantities.

The single most precise measurement of $\ssteff$ comes from the
precise measurement of the left-right polarisation asymmetry at
SLD. This asymmetry is defined simply as 
\begin{equation}
\alr = \frac{(N_L - N_R)}{(N_L + N_R)}\frac{1}{\langle P_e \rangle}
\end{equation}
where $N_L$($N_R$) are the numbers of observed \PZz\ events for
left and right handed beam polarisations, and $\langle P_e \rangle$ is
the mean electron beam polarisation.
Knowledge of the mean beam polarisation is therefore a major
experimental challenge of this measurement, and it is
measured with three different techniques at SLD. 
The final result obtained\cite{alr} is $\alrz=0.1514\pm0.0022$, which
can be
combined with other SLD asymmetry measurements\cite{sldasy} to give 
$\ssteffl=0.23098\pm0.00026$.

\begin{figure}[htb]
\begin{center}
\mbox{\epsfysize=6cm\epsfbox{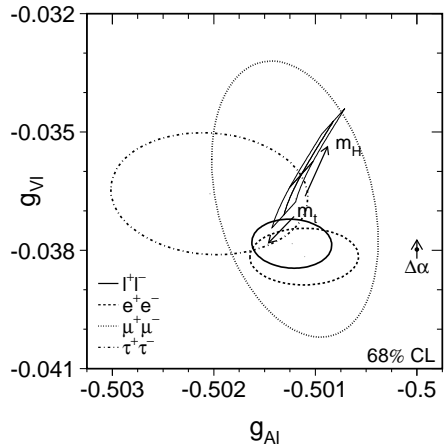}}
\end{center}
\caption{Measured couplings of the \PZz\ to charged leptons, compared
  with the Standard Model prediction. This prediction is given for a
  range of Higgs masses between 114 and 1000 GeV,
  and top mass of 174.3$\pm$5.1~GeV. The arrows indicate the
  directions of increasing mass.}
\label{fig:lepton}
\end{figure}
Combining together the leptonic asymmetry measurements gives an
overall value
of $\Aparl=0.1501\pm0.0016$.
This may be further combined with \PZz\ leptonic decay widths to
extract the individual couplings of the
\PZz\ to the individual lepton species. The results are shown in
figure~\ref{fig:lepton}, both without and with the assumption of
lepton universality. The data are seen to be quite consistent with
lepton universality in the neutral current. It is also
evident that there is a strong sensitivity to the effects of
electroweak radiative corrections, beyond the well-known effects
of photon radiation, which alone would give the prediction shown at
the right-hand side of the plot, labelled $\Delta\alpha$.
This illustrates the sensitivity of the precise electroweak data to
the mass
of heavy particles such as the top quark and Higgs boson via loop
corrections, when the Standard Model structure of radiative
corrections is assumed.

\begin{figure}[htb]
\begin{center}
\mbox{\epsfysize=7cm\epsfbox{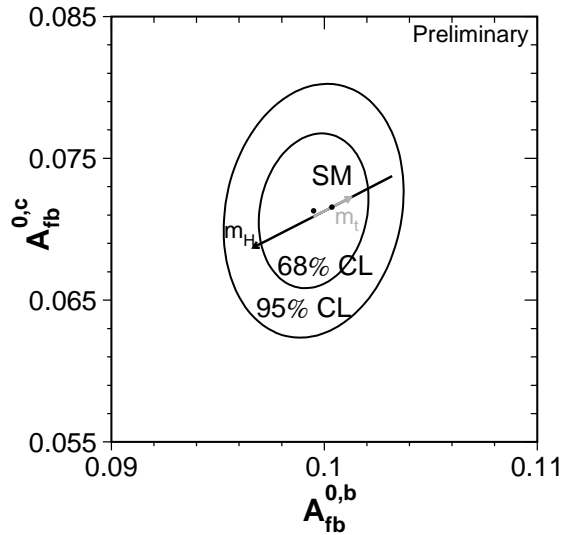}}
\end{center}
\caption{Measured heavy quark forward-backward asymmetries compared
  with the Standard Model predictions. The arrows on the prediction
  show the effect of varying the Higgs mass between 114 and 1000 GeV,
  and the top mass over 174.3$\pm$5.1~GeV. }
\label{fig:afbq}
\end{figure}

In addition to those for charged leptons, forward-backward asymmetries
are also measurable for b and c quarks. Excellent flavour tagging is
needed, as for the measurements of $R_b$ and $R_c$. 
In addition, charge tags are needed to separate
quark-initiated jets from those initiated by antiquarks. This is
typically provided by multivariate discriminants whose performance is
calibrated with data, or from semileptonic decay
tags.
The measured values\cite{afbq} of $\afbb$ and $\afbc$ are shown in
figure~\ref{fig:afbq} after combination between experiments.

It is interesting to note that, although the heavy quark
forward-backward asymmetries measure the combination $\Apare\Aparf$
(f=b,c), the main sensitivity to the electroweak mixing angle
$\ssteff$ arises from $\Apare$, so that the measurements
effectively probe $\ssteffl$. $\Aparb$ and $\Aparc$ are relatively
precisely predicted by the Standard Model, with little dependence on
the Higgs or top quark masses.

\begin{figure}[htb]
\begin{center}
\mbox{\epsfysize=12cm\epsfbox{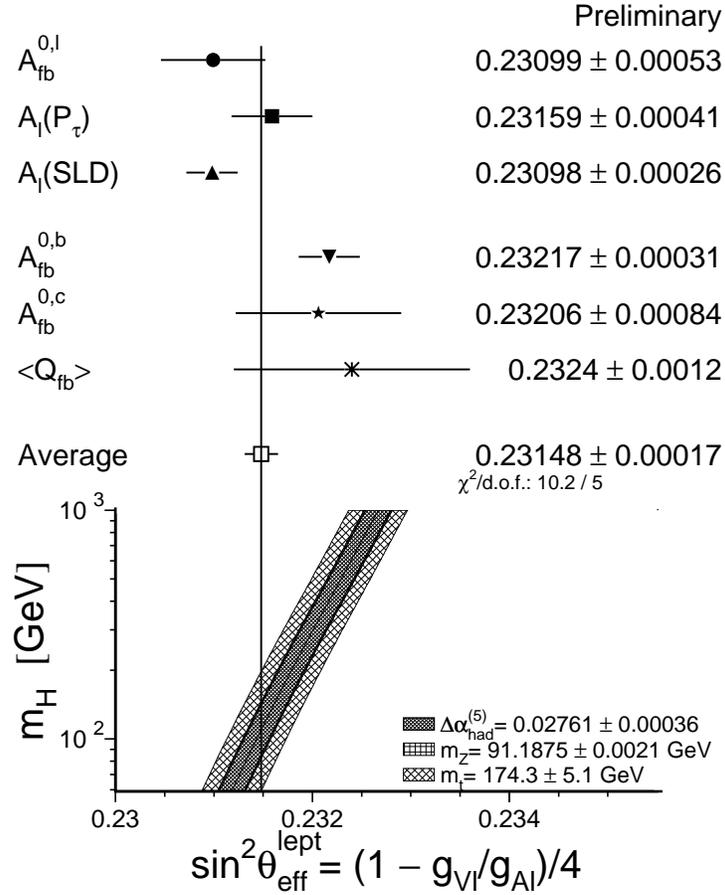}}
\end{center}
\caption{$\ssteffl$ as derived from various asymmetry
  measurements. The Standard Model prediction (bottom) is shown for a
  range of Higgs masses.}
\label{fig:ssteff}
\end{figure}

A comparison of the $\ssteffl$ values extracted from the different
asymmetry measurements is shown in figure~\ref{fig:ssteff}.
A posteriori, it is observed that the two most precise $\ssteffl$
measurements agree only at the 2.9$\sigma$ level.
However, this is an old problem: the discrepancy has been around 
3$\sigma$ for the last six years, even though the measurement errors
have improved by a factor of 1.5.
In the context of the Standard Model, the results can alternatively be
summarised as saying that $\afbb$ prefers a Higgs mass
$\Mh\sim$~400~GeV --- unlike most other observables which prefer a low
$\Mh$, close to, or even below, the current direct search limits.

\section{LEP-2: Fermion and W Pair Production}

\subsection{Fermion-Pair Production at High Energy}

\begin{figure}[htb]
\begin{center}
\mbox{\epsfysize=7cm\epsfbox{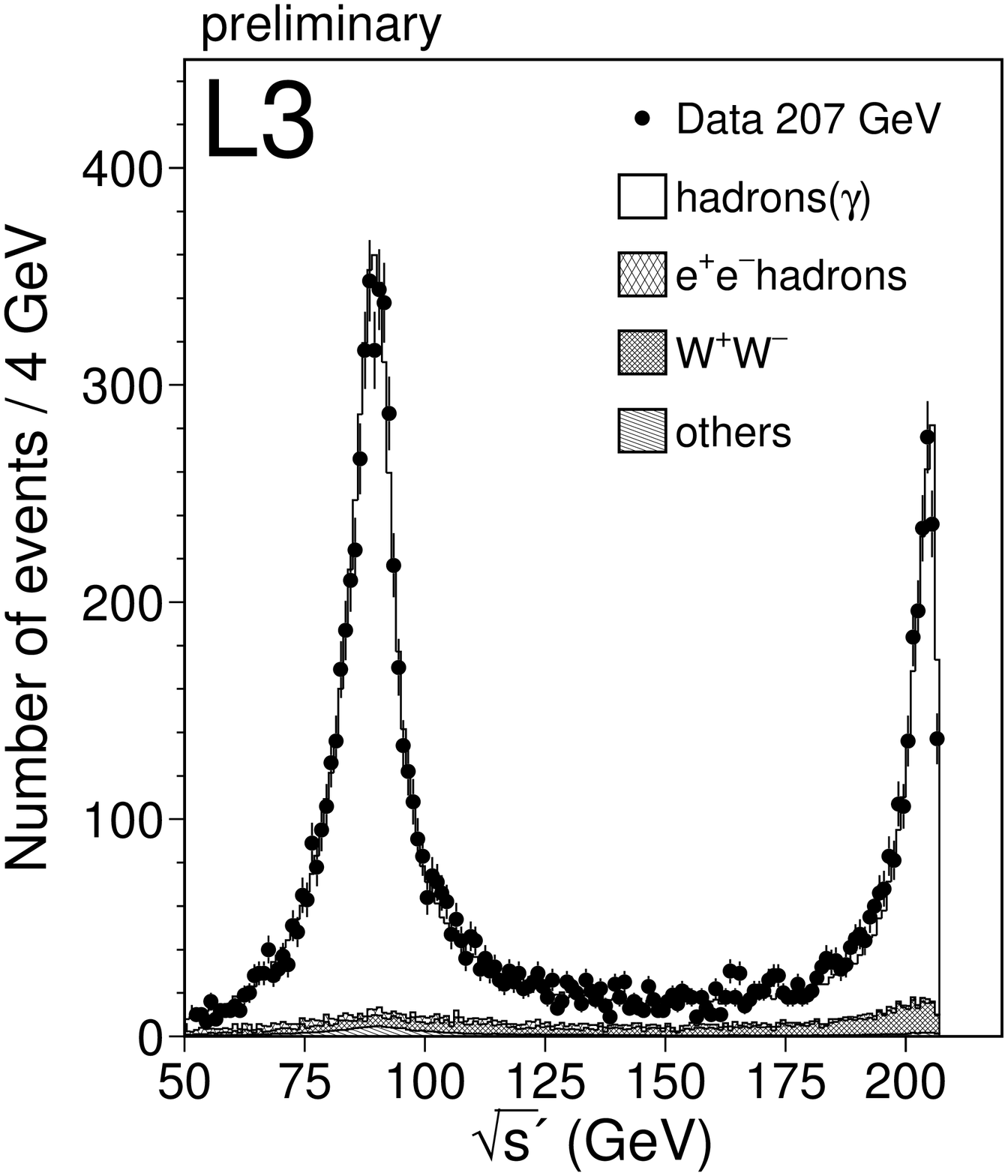}\epsfysize=7.5cm\epsfbox{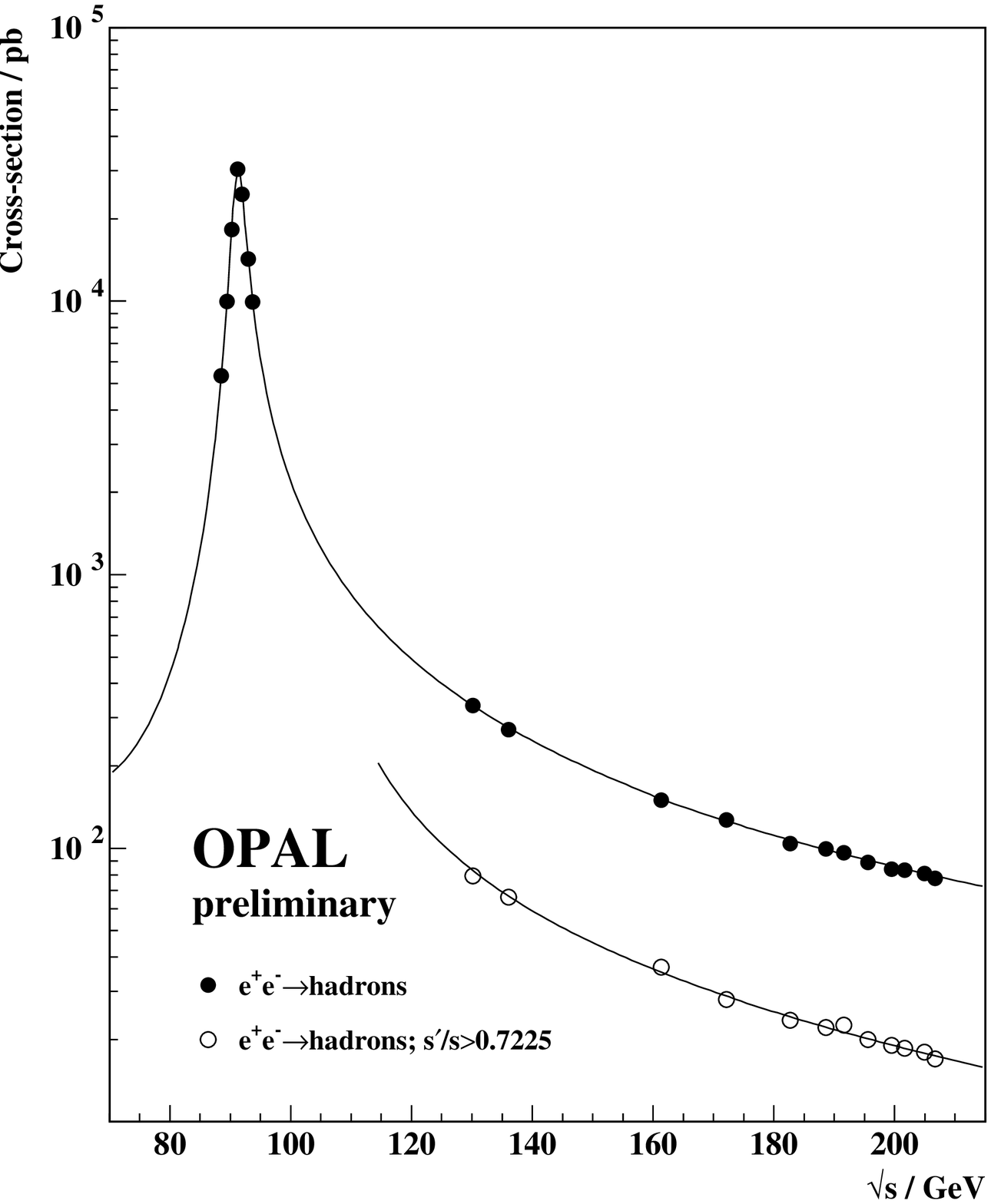}}
\end{center}
\caption{Fermion-pair production at LEP-2. Left: $\sqrt{s^{\prime}}$
  distribution for $\qqbar$ events at 207 GeV centre-of-mass energy;
  Right: $\qqbar$ cross-section, inclusive and non-radiative.}
\label{fig:fflep2}
\end{figure}
At LEP-2 centre-of-mass energies (130-209 GeV) fermion-pair events
very largely divide into two characteristic populations according
to the invariant mass ($\sqrt{s^{\prime}}$) of the final-state fermions.
This is illustrated in figure~\ref{fig:fflep2}(left): some events have
$\sqrt{s^{\prime}}$ close to the full centre-of-mass energy
(``non-radiative'') while others have one or more hard initial-state
radiation photons which have lowered $\sqrt{s^{\prime}}$ to around
$\Mz$ (``radiative return'').
The sizeable contribution of radiative return events is illustrated by
figure~\ref{fig:fflep2}(right).

\begin{figure}[htb]
\begin{center}
\mbox{\epsfysize=8cm\epsfbox{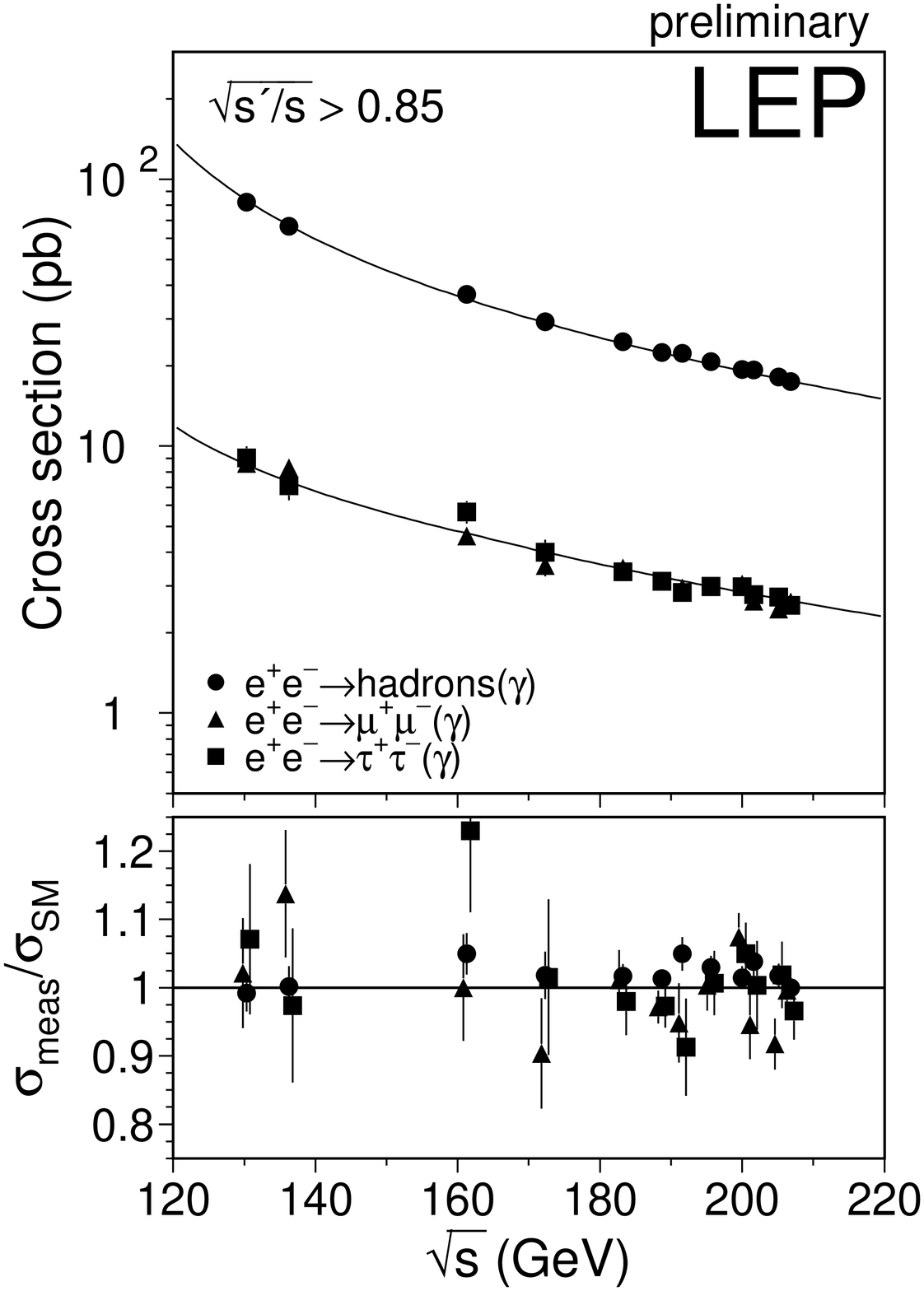}\epsfysize=8cm\epsfbox{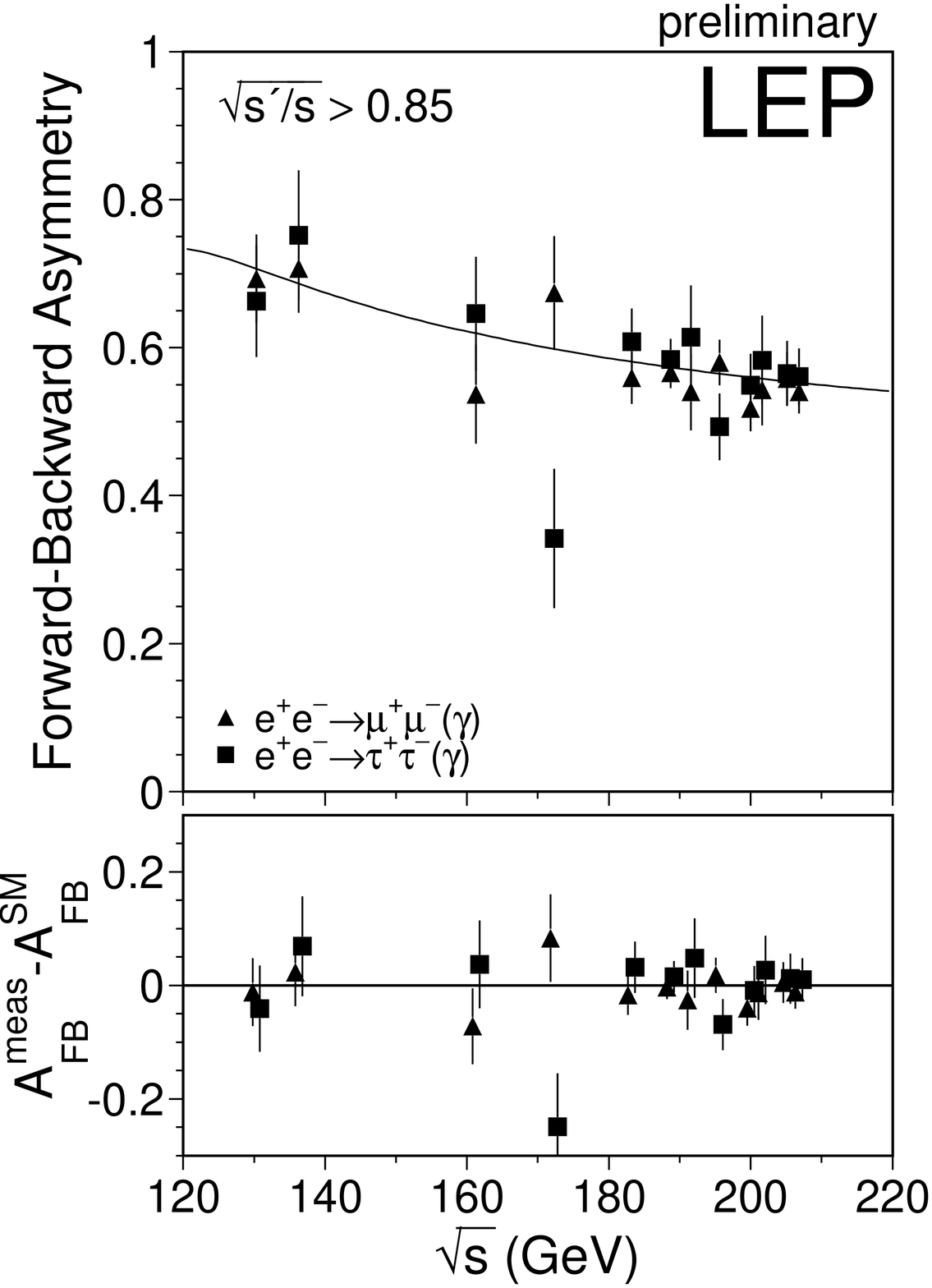}}
\end{center}
\caption{Measured non-radiative fermion-pair cross-sections (left) and
  asymmetries (right) at LEP-2 energies, and compared to Standard
  Model predictions.}
\label{fig:ffmeas}
\end{figure}

The properties of non-radiative events thus probe the full
centre-of-mass energy of the colliding beams, in contrast to the
radiative return events which should be typical \PZz\ decays. 
It is therefore particularly interesting to use the 
non-radiative cross-sections and asymmetries to probe the quality of 
the Standard Model predictions:
deviations could indicate effects of new physics at higher mass
scales.
LEP combined cross-section and forward-backward asymmetry
measurements\cite{fflep2}
are shown in figure~\ref{fig:ffmeas}. 
The Standard Model indeed continues to describe fermion-pair
production at LEP-2 energies.

\subsection{Production and Decays of W Bosons}

Most of the key physics goals of LEP-2 are based around the production
of pairs of W bosons.
Since the events contain two W's, the topologies fall into three quite
distinct categories, according to whether both, one or neither W
decays hadronically. 
In approximately 46\% of WW events both W's decay to quarks, giving a
typically four-jet final state, in a similar fraction (44\%) one
decays hadronically and one leptonically, giving two jets, a charged
lepton and missing momentum from the neutrino, and finally 10\% of WW
events have two charged leptons with large acoplanarity arising from
the two unobserved neutrinos.

\begin{figure}[htb]
\begin{center}
\mbox{\epsfysize=7cm\epsfbox{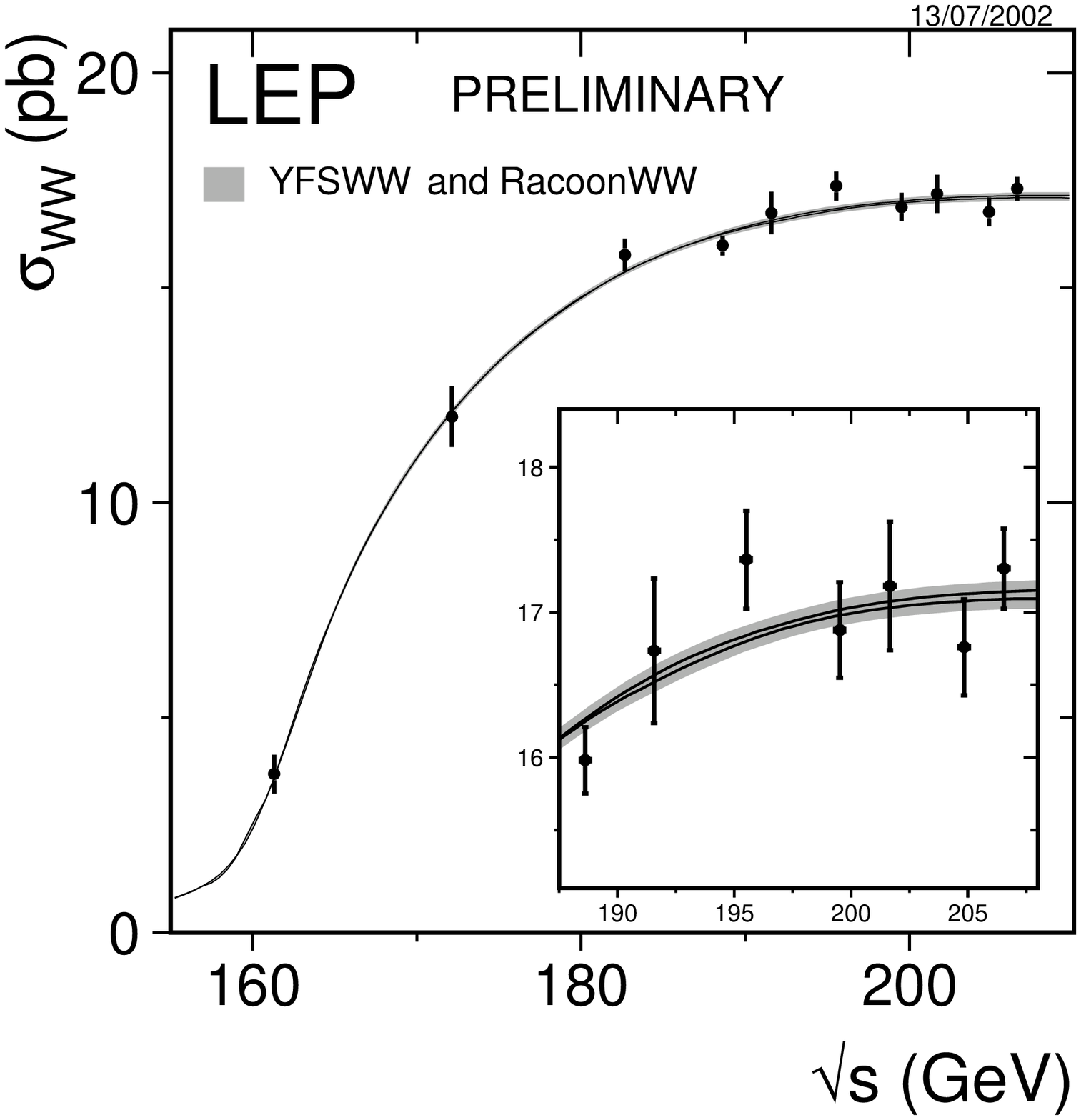}\epsfysize=7cm\epsfbox{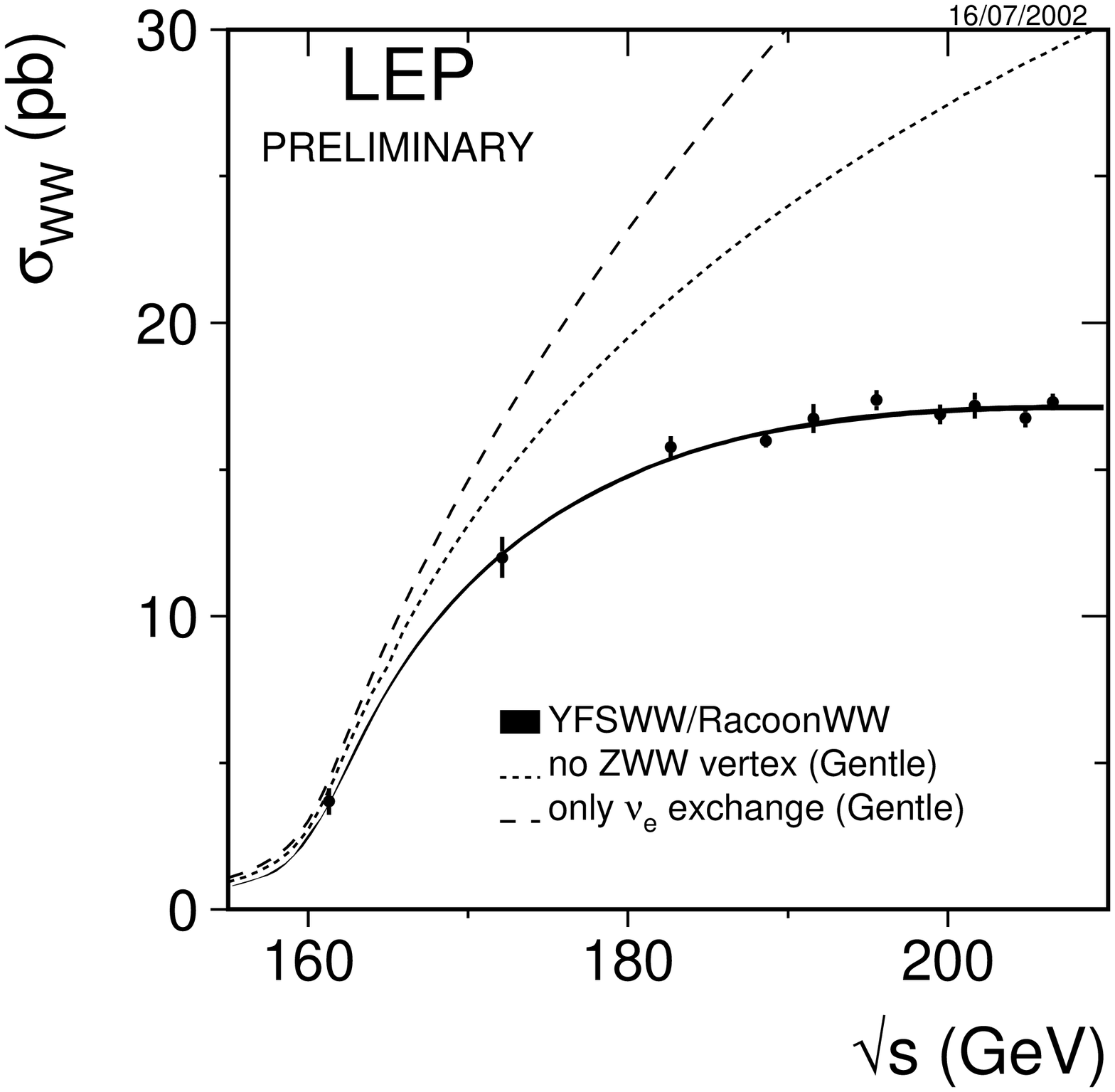}}
\end{center}
\caption{Measured WW cross-sections at LEP-2, (left) compared to
  predictions including O($\alpha$) effects, and
  (right) showing the effect if diagrams containing triple
  vector-boson vertices are omitted.}
\label{fig:sigww}
\end{figure}

The selection of WW events is by now well established in the four LEP
experiments\cite{wsel}: typical selection efficiencies and purities 
are 80-90\%.
The measured W-pair cross-section is shown in
figure~\ref{fig:sigww}. With a total LEP-2 integrated luminosity of
around 700\,\ipb\ per experiment, around 12000 WW events are observed
by each.

Recently, substantial theoretical progress has been made incorporating
O($\alpha$) corrections into the predictions for W-pair
production. This is illustrated in the inset of
figure~\ref{fig:sigww}, where the measurements are compared with
predictions of the YFSWW\cite{yfsww} and RacoonWW\cite{racoon} 
Monte Carlo programs. 
These calculations, including the main O($\alpha$) effects, describe
the data well, but lie approximately 2\% below predictions without
these
corrections.

\subsection{Gauge Structure of the Standard Model}

\begin{figure}[htb]
\begin{center}
\mbox{\epsfysize=2.2cm\epsfbox{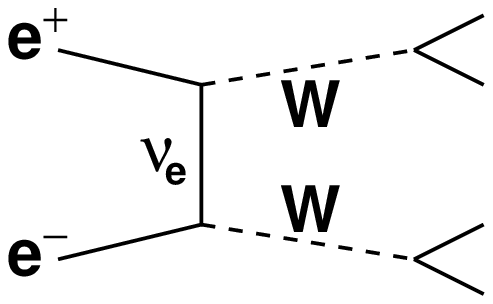}\hspace*{5mm}\epsfysize=2.2cm\epsfbox{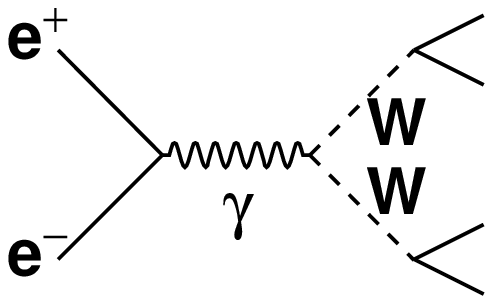}
\hspace*{5mm}\epsfysize=2.2cm\epsfbox{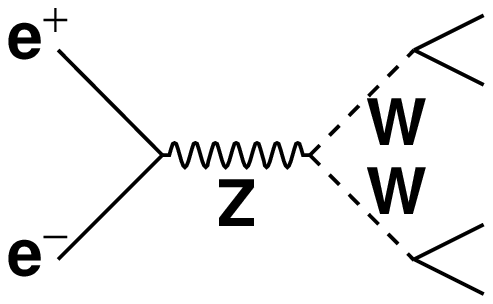}}
\end{center}
\caption{Doubly-resonant W-pair production diagrams.}
\label{fig:wwdiag}
\end{figure}

Another important feature of W-pair production at LEP-2 is the
substantial effect of diagrams containing vector-boson self-couplings,
as shown in figure~\ref{fig:wwdiag} along with the neutrino exchange
diagram. 
The size of the contribution from these diagrams is shown in
figure~\ref{fig:sigww}: with only neutrino exchange the cross-section
would be much higher and would eventually violate unitarity at higher
centre-of-mass energies.

Sensitivity to the properties of the triple vector boson vertices
arises not only through the total WW cross-section, but also through
the differential cross-section as a function of the scattering polar
angle, and through the W polarisation as a function of angle. The
polarisation can be probed using the W decays as helicity analysers. 
The analyses\cite{tgc} of triple-gauge couplings at LEP-2 make use of
all these
properties to extract the most information about vertex coupling
factors: conventionally the main parameters studied are known as
$\kappa_{\gamma}$, $\lambda$ and $g_Z^1$. In the Standard Model these
are respectively 1, 0 and 1. In extracting these parameters from the
data the effects of O($\alpha$) corrections are significant, and so
these are now included in the analyses.
The current, preliminary, LEP averaged results
\begin{eqnarray}
\kappa_{\gamma} & = & \phantom{-}0.943\pm0.055 \\
\lambda & = & -0.020\pm0.024 \\
g_Z^1 & = & \phantom{-}0.998\pm_{0.025}^{0.023}
\end{eqnarray}
are consistent with these predictions, thus directly demonstrating
the gauge structure of the Standard Model in the vector boson
self-couplings.

\subsection{Measurement of the W Mass}

\begin{figure}[htb]
\begin{center}
\mbox{\epsfysize=10cm\epsfbox{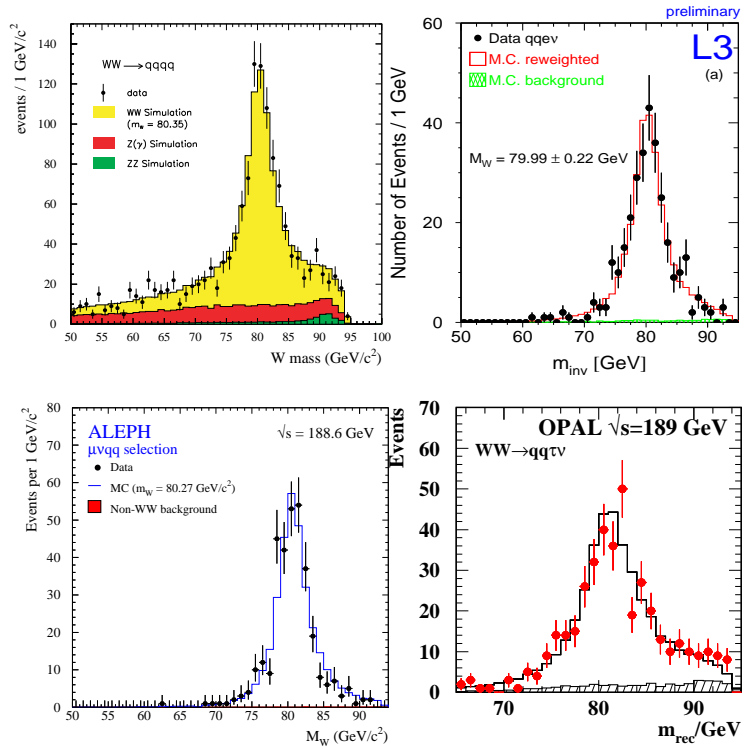}}
\end{center}
\caption{Reconstructed W mass distributions in the different WW decay
  channels.}
\label{fig:mwdist}
\end{figure}

The copious production of W pair events at LEP-2 allows the
measurement of the W mass. Experimentally this employs the 
reconstructed directions and energies of the primary W decay
products, using jet directions and energies to approximate those of
the primary quarks, measuring charged lepton momenta, and deducing
neutrino momenta from the missing momentum in the event. 
Information comes primarily from events where at least one W decays
hadronically -- the double leptonic decays having too few observables
to extract much information on $\Mw$. 
For the $\PWp\PWm\to\qqqq$ and $\PWp\PWm\to\qqln$ events, a kinematic
fit is made to improve the W mass resolution event by event: this fit
imposes the constraints that the total energy and momentum of the W
decay products equal that of the colliding beams, and (usually) that
the two decaying W's have the same mass.
Whilst these constraints are not physically exact, due to
initial-state radiation and the finite W width, these can be
modelled accurately in Monte Carlo simulation and so corrected
for. The corrections are normally included in the fitting procedure
by comparing data directly with simulated Monte Carlo.
Typical reconstructed mass distributions are shown in
figure~\ref{fig:mwdist}.
The width of the distributions has a substantial contribution from
the W width, and so fits may be made both for $\Mw$ alone, and for
$\Mw$ and $\Gw$ simultaneously.

\begin{figure}[htb]
\begin{center}
\mbox{\epsfysize=6cm\epsfbox{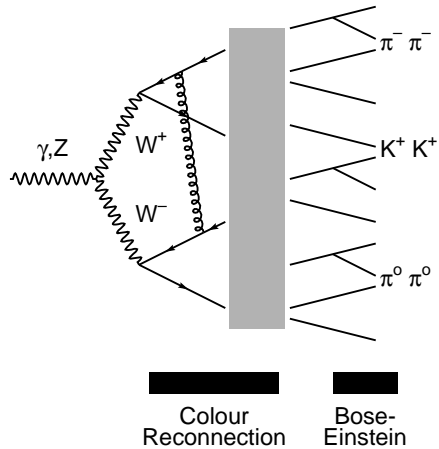}}
\end{center}
\caption{Schematic of final-state interaction models in doubly
  hadronic W-pair decays.}
\label{fig:fsi}
\end{figure}
With the full LEP-2 statistics, systematic errors are significant
compared to the statistical errors, especially in the
$\PWp\PWm\to\qqqq$ channel, which is now systematics dominated.
In the absence of systematic errors, the statistical precision of the
$\qqqq$ and $\qqln$ channels is similar: the effect of the large
systematics in the $\qqqq$ channel is to deweight very substantially 
the contribution of
this channel in the average.
Much the most serious difficulty in the double hadronic decay channel
arises from the fact that the two W's decay with a separation much
less than a typical hadron size. 
It is thus very possible that the two W decays interact with each
other in
the hadronisation process -- the effect has been studied analytically
in the perturbative phase and found to be small\cite{crpert}, 
but such a conclusion
cannot be drawn for the non-perturbative region.
Models are therefore necessary to consider such effects, and
these are divided into two physical types,
``colour-reconnection'' and Bose-Einstein correlations, as indicated
schematically in figure~\ref{fig:fsi}.

The effect of Bose-Einstein correlations between the hadronic decay
products of different W's has been studied\cite{bec} by the LEP
experiments, which currently indicate that the effect on $\Mw$ in the
hadronic channel is quite small. 
Colour-reconnection (CR) models are also being studied in
detail\cite{cr}, 
but here
the effects are found to be large, and hard to control.
Since colour reconnection may be thought of as the exchange of
(multiple) soft gluons, it is expected primarily to affect soft
particle production.
Substantial effort has been invested into understanding the relative
effects of CR on the W mass measured by different experiments: the
effects are found to be very similar, as might be expected.
Several studies look at the distributions of soft particles between
jets, where the effects of colour reconnection might be most visible: 
sensitivity is found to some models, such as the
``SK-I'' model implemented in JETSET\cite{ski}, 
but not to others, such as the
``AR-2'' CR model implemented in Ariadne\cite{arcr}.
Overall, the effect of these studies has thus been to establish rather
well the effect of the different models on $\Mw$, and to eliminate
some of the most extreme cases, but the remaining systematic
uncertainty on the W mass from the $\qqqq$ channel from this source is
still estimated to be as much as 90~MeV\cite{cr}.

\begin{figure}[htb]
\begin{center}
\mbox{\epsfysize=7.5cm\epsfbox{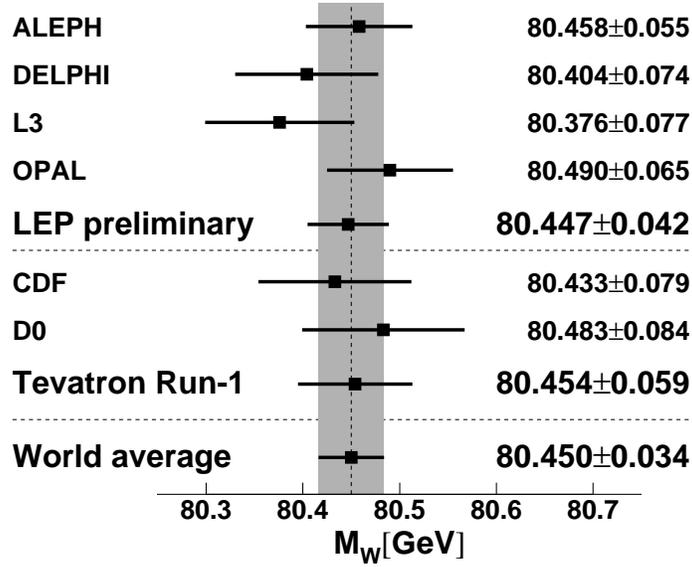}}
\end{center}
\caption{W mass measurements from LEP and the Tevatron.}
\label{fig:mw}
\end{figure}
Including these and all other systematic errors, the W mass values
obtained by the four LEP experiments\cite{mwlep}, 
and their combination, is shown
in figure~\ref{fig:mw}. The weight of the $\qqqq$ channel in the LEP
average is just 9\%. 
The LEP results are compared to those from the Tevatron\cite{mwtev}: 
the
individual measurements from the LEP experiments are of comparable
precision to those from the Tevatron. 
Since the extraction methods are so different the LEP and Tevatron
$\Mw$ results are quite uncorrelated, so combining the
measurements results in a substantial improvement, giving a world
average W mass precision of 34~MeV.

\begin{figure}[htb]
\begin{center}
\mbox{\epsfysize=7.5cm\epsfbox{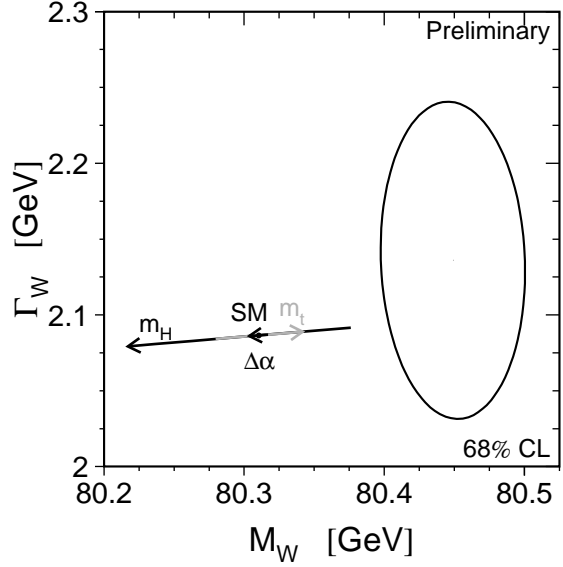}}
\end{center}
\caption{W mass and width results, compared to the Standard Model
  expectation.
This prediction is given for a
  range of Higgs masses between 114 and 1000 GeV,
  and top mass of 174.3$\pm$5.1~GeV. The arrows indicate the
  directions of increasing mass.}
\label{fig:mwgw}
\end{figure}

The LEP experiments' results for the W width\cite{mwlep} 
give a preliminary value of
$\Gw=2.150\pm0.091$~GeV when combined, compared to
a value of $\Gw=2.115\pm0.105$~GeV obtained from the Tevatron
measurements using the high-$p_T$ tail of the lepton momentum
distribution\cite{gwtev}.
The combined $\Gw$ vs $\Mw$ contours are shown in
figure~\ref{fig:mwgw}: the W width is in good agreement with the
Standard Model expectation; the W mass is quite sensitive to the Higgs
mass in the Standard Model framework, and favours a low Higgs mass.

\section{Global Electroweak Tests}

The various electroweak measurements discussed in the preceding
sections can be combined to provide global tests of the description of
the precise electroweak data by the Standard Model.
In addition, results from NuTeV\cite{nutev} and atomic parity
violation\cite{apv} are included.
The Standard Model predictions are provided by the
ZFITTER\cite{zfitter} and TOPAZ0\cite{topaz}
electroweak libraries.
Typically in the fit, parameters such as $\Mz$, $\alpha_s(\Mz^2)$, and
$\alpha_{em}(\Mz^2)$ are allowed to vary, but are strongly constrained
by measurements at the Z lineshape ($\Mz$, $\alpha_s$)
or lower energies ($\alpha_{em}$).
Of most interest are the parameters $\Mw$,$\Mtop$ and $\Mh$, all three
of which may be predicted from fits to data excluding any direct
measurements. 
In the case of the W and the top these may be compared with the direct
measurements, and in the case of $\Mh$ the predictions can be compared
with current experimental search limits.

\begin{figure}[htb]
\begin{center}
\mbox{\epsfysize=8cm\epsfbox{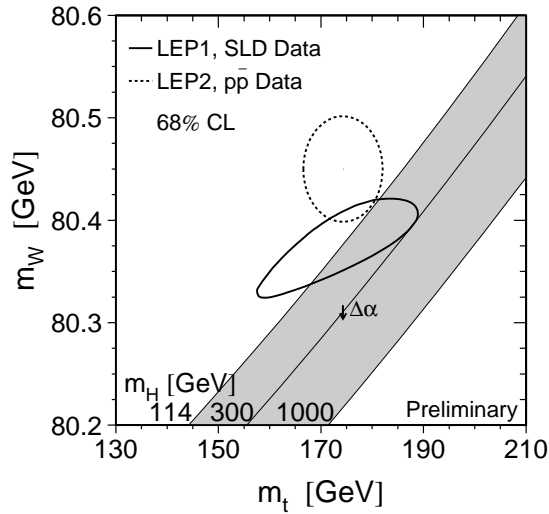}}
\end{center}
\caption{Predicted $\Mw$ and $\Mtop$: from the fit to Z pole and 
  lower energy
  measurements (solid contour); from direct measurements (dashed); and
  using Standard Model inter-relations (diagonal bands) as a function
  of Higgs mass.}
\label{fig:mtmw}
\end{figure}
The results of a fit including all electroweak data except the direct
W and top mass measurements is shown in figure~\ref{fig:mtmw}, with
results shown by the solid contour. 
The predictions are consistent with the direct measurements of these
masses, shown by the dashed ellipse, demonstrating that the Standard
Model fit can correctly predict the masses of heavy particles, in the
case of top via radiative loop corrections. 
The ellipses are compared also with the Standard Model prediction
of the relationship between $\Mw$ and $\Mtop$ as a function of $\Mh$
(diagonal bands): it is seen that both the precise lower energy data,
and the direct measurements of the masses, favour a low Higgs mass in
the Standard Model framework.

\begin{figure}[htp]
\begin{center}
\mbox{\epsfysize=14cm\epsfbox{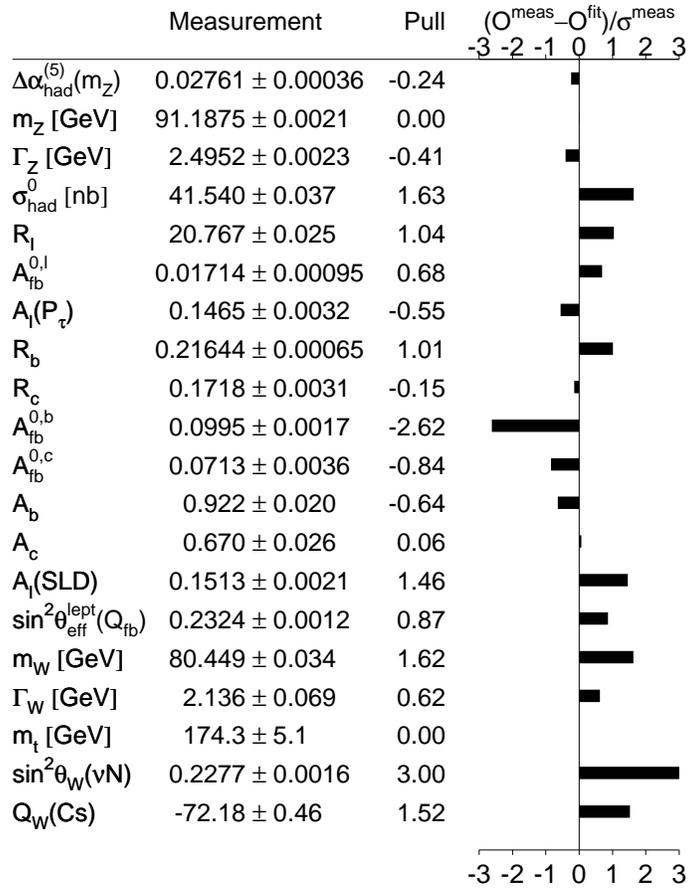}}
\end{center}
\caption{Measured and fitted electroweak parameters.}
\label{fig:pulls}
\end{figure}

A full electroweak fit is made including all the precise electroweak
data, as indicated in figure~\ref{fig:pulls}. The ``pull'' indicated
is defined as the measured value of the observable minus the best fit
value divided by the measurement error only.
The overall $\chi^2$ of the fit is 29.7 for 15 degrees of freedom,
corresponding to a 1.3\% fit probability. 
This rather high $\chi^2$ has the largest contribution from the new
NuTeV
measurement of $\sstw$ in neutrino-nucleon scattering\cite{nutev}.
Without this measurement, the fit $\chi^2$ would be 11\%.
The next largest contribution comes from the b quark forward-backward
asymmetry measurement, as noted earlier.
It is worth noting that although the $\chi^2$ for the fit is
increased substantially by inclusion of the new NuTeV result, the
electroweak parameters, particularly $\Mh$, extracted from the fit are
little affected by whether this result is included or not.
It is therefore of interest to go on to see what the Standard Model
fit says about $\Mh$.

\begin{figure}[htb]
\begin{center}
\mbox{\epsfysize=10cm\epsfbox{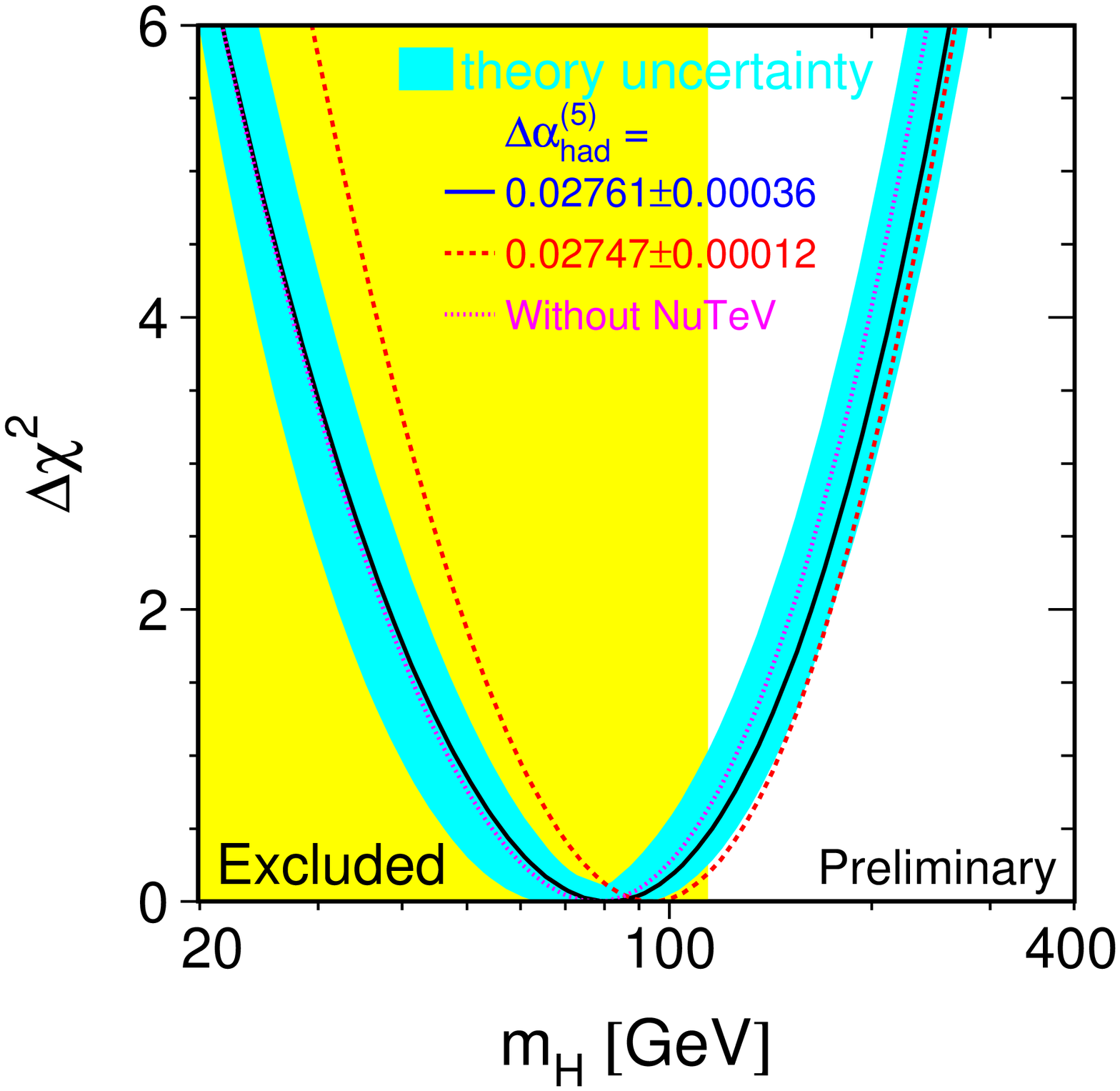}}
\end{center}
\caption{Chi-squared of the Standard Model fit for different assumed
  Higgs boson masses. See text for details.}
\label{fig:blueband}
\end{figure}
The $\chi^2$ of the full Standard Model electroweak fit to all results
is shown in figure~\ref{fig:blueband} as a function of the Higgs
mass. The shaded band around the central curve shows the effect of
higher-order theoretical uncertainties, evaluated by
varying ZFITTER/TOPAZ0 options and including an estimate of the effect
of a partial inclusion of two-loop corrections. The lighter shaded
region to the left of the plot indicates the region experimentally
excluded by direct searches, as discussed in the next section.
The Higgs mass obtained from the fit is:
\begin{equation}
\Mh = 81_{-33}^{+52} \mathrm{~GeV}
\end{equation}
corresponding to $\Mh < 193$~GeV at 95\% CL. 
This result includes preliminary results, and is valid only in the
framework of the Standard Model.

\section{The Search for the Higgs Boson of the Standard Model}

Direct searches for production of Higgs bosons were carried out
each time the centre-of-mass energy of LEP was raised. The final
results\cite{higgs} are dominated by the highest energy data, which
are reviewed briefly.
Since the couplings of the Higgs boson are completely fixed in the
Standard Model for any given Higgs mass, $\Mh$, this is the only
free parameter in these searches (this is relaxed if supersymmetric
extensions to the Standard Model are considered, but these are beyond
the scope of this paper).
The main production process at LEP-2 energies would be the so-called
``Higgs-strahlung'' process, $\Pep\Pem\to\PZz\mathrm{H^0}$. For Higgs
masses in the relevant range 80~$<\Mh<$~120\,GeV, the main decay mode
would be H$\to\bbbar$, so that the key experimental techniques are
high performance b tagging, and an excellent mass resolution to
separate any possible signal from the irreducible background from
$\PZz\PZz$ 
production.

It was therefore intriguing when ALEPH reported\cite{haleph} an excess
of three high-mass 
events in the $\qqbar\bbbar$ channel in September 2000, just before
the planned end of LEP data-taking. The three events were consistent
in their kinematics with production and decay of a Higgs of mass
115~GeV, although the rate was rather higher than expected. A long,
and at times heated, debate followed, as a result of
which a brief one-month extension of the LEP run was granted.

Although the original events from ALEPH remain, the full analyses of
the data of the other three experiments\cite{higgs} show no excess of
events that might confirm the hypothesis suggested by the ALEPH data, 
and no significant excess is observed in the final combined
sample.
Instead a lower limit\cite{lephiggs} can be placed on the mass of a
Standard Model
Higgs boson of
\begin{equation}
\Mh>114.4\mathrm{~GeV~(95\%~CL)}
\end{equation}
slightly below the expected limit of 115.3~GeV.

\section{Conclusions}

A wealth of electroweak measurements have been collected by LEP and
SLD in the last thirteen years, and only a brief overview of them
could be
given here. They are complemented in key places by measurements
from the Tevatron.

To select just a few from the very large range of highlights, these
data have:
\begin{itemize}
\item shown there are three light neutrino species
\item demonstrated radiative loop corrections
\item predicted the top quark mass
\item verified Standard Model triple gauge couplings
\item put many strong constraints on physics beyond the SM
\item indicated where to look for the Standard Model Higgs -- although
  direct observation was not possible
\end{itemize}
In total, therefore, LEP and SLD have provided a huge step forward for
the Standard Model,
but the final elucidation of the Higgs sector and the mechanism of
electroweak symmetry breaking awaits future, higher energy,
experiments.
   
\section*{Acknowledgments}

Much credit for the LEP combined results and plots in this note goes
to the LEP/SLD electroweak working group together with its various 
subgroups, in addition to the members of the five experiments. 
In particular I want to thank Martin Gr\"unewald for his prompt
preparation of combined results, and willingness always to explain.
Thanks also to Arno
Straessner for figure~\ref{fig:fsi}.


\begin{thebibliography}{10}

\bibitem{lineshape}
ALEPH Collaboration, R.~Barate \etal, Eur. Phys. J. {\bf C 14} 1 (2000);\\
DELPHI Collaboration, P.~Abreu \etal, Eur. Phys. J. {\bf C 16} 371 (2000);\\
L3 Collaboration, M.~Acciarri \etal, Eur. Phys. J. {\bf C16} 1-40 (2000);\\
OPAL Collaboration, G.~Abbiendi \etal, Eur. Phys. J. {\bf C19} 587 (2001).
\bibitem{leplineshape}
The LEP~Collaborations  ALEPH, DELPHI, L3 and OPAL 
and the LEP Electroweak working group,
{\it Combination procedure for the precise determination of Z boson 
parameters from results of the LEP experiments}, CERN--EP-2000-153, 
hep-ex/0101027.
\bibitem{rbrc}
ALEPH Collaboration, R.~Barate \etal,
Phys. Lett. {\bf B 401} 150 (1997);\\
ALEPH Collaboration, R.~Barate \etal,
Phys. Lett. {\bf B 401} 163 (1997);\\
ALEPH Collaboration, R.~Barate {\em et~al.}, 
Eur. Phys. J. {\bf C4} 557 (1998);\\
ALEPH Collaboration, R. Barate \etal,
Eur. Phys. J. {\bf C16} 597 (2000);\\
DELPHI Collaboration, P.~Abreu \etal, Eur. Phys. J. {\bf C10} 415 (1999);\\
DELPHI Collaboration, P.~Abreu \etal, Eur. Phys. J. {\bf C12} 209 (2000);\\
DELPHI Collaboration, P.~Abreu \etal, Eur. Phys. J. {\bf C12} 225 (2000);\\
L3 Collaboration, M. Acciarri \etal,
Eur. Phys. J. {\bf C13} 47 (2000); \\
OPAL Collaboration, G.~Alexander \etal, Z.~Phys.~{\bf C72} 1 (1996);\\
OPAL Collaboration,  K.~Ackerstaff \etal, 
Eur. Phys. J. {\bf C1} 439 (1998);\\
OPAL Collaboration, G. Abbiendi \etal, 
Eur. Phys. J. {\bf C8} 217 (1999); \\
SLD Collaboration, K. Abe  \etal, Phys. Rev. Lett. {\bf 80} 660 (1998);\\
SLD Collaboration, SLAC--PUB--7880,
contributed paper to ICHEP98;\\
N. De Groot, ``Electroweak results from SLD'',
hep-ex/0105058;\\
V. Serbo, presented at Int. Europhys. Conf. High Energy Phys.,
July 2001, Budapest, Hungary.
\bibitem{rbcrisis}
See, for example, D.G.~Charlton, BHAM-HEP-95-01, invited talk at
EPS-HEP95, Brussels, Belgium (1995).
\bibitem{alr}
SLD Collaboration, K.~Abe \etal, Phys. Rev. Lett. {\bf 84} 5945 (2000).
\bibitem{sldasy}
SLD Collaboration, K.~Abe \etal,
Phys. Rev. Lett. {\bf 86}, 1162 (2001)
\bibitem{afbq}
ALEPH Collaboration, A.~Heister \etal,
Eur. Phys.\ J. {\bf C24} 177 (2002);\\
ALEPH Collaboration, A.~Heister \etal,
Eur. Phys. J. {\bf C 22} 201 (2001);\\
ALEPH Collaboration, R.~Barate \etal, Phys. Lett. {\bf B434} 415 (1998);\\
DELPHI Collaboration, P.~Abreu \etal, Z.~Phys {\bf C65} 569 (1995);\\
DELPHI Collaboration, DELPHI 2002-028-CONF-562 (2002);\\ 
DELPHI Collaboration, DELPHI 2002-029-CONF-563 (2002);\\
DELPHI Collaboration, P.~Abreu \etal, Eur. Phys. J. {\bf C9} 367 (1999);\\
DELPHI Collaboration, P.~Abreu \etal, Eur. Phys. J. {\bf C10} 219 (1999);\\
L3 Collaboration, O.~Adriani \etal, Phys.~Lett. {\bf B292 } 454 (1992);\\
L3 Collaboration, L3 Note 1624 (1994);\\
L3 Collaboration, M. Acciarri \etal, Phys. Lett. {\bf B448} 152 (1999);\\
L3 Collaboration, M. Acciarri \etal, Phys. Lett. {\bf B439} 225 (1998);\\
OPAL Collaboration,
G.~Alexander \etal, Z.~Phys. {\bf C70 } 357 (1996);\\
OPAL Collaboration, R.~Akers \etal, OPAL Physics Note PN226 (1996); \\ 
OPAL Collaboration, R.~Akers \etal, OPAL Physics Note PN284 (1997); \\
OPAL Collaboration, K.~Ackerstaff \etal, Z.~Phys. {\bf C75} 385 (1997);\\
OPAL Collaboration, G.~Alexander \etal, Z.~Phys. {\bf C73} 379 (1996).
\bibitem{fflep2}
LEP EW WG $\ffbar$ subgroup, C.\,Geweniger \etal, Note LEP2FF/02-03 (2002),
and references therein.
\bibitem{wsel}
ALEPH Collaboration, R.~Barate \etal, Phys. Lett. {\bf B484} 205 (2000);\\
DELPHI Collaboration, P.~Abreu \etal, Phys. Lett. J. {\bf B479} 89 (2000);\\
L3 Collaboration, M. Acciarri \etal, Phys. Lett. {\bf B496} 19 (2000);\\
OPAL Collaboration, G.~Abbiendi \etal, Phys. Lett. {\bf B493} 249 (2000).
\bibitem{yfsww}
S.~Jadach, W.~Placzek, M.~Skrzypek, B.~F.~Ward and Z.~Was,
Comput.\ Phys.\ Commun.\  {\bf 140} 432 (2001).
\bibitem{racoon}
A.\,Denner, S.\,Dittmaier, M.\,Roth and D.\,Wackeroth, Nucl. Phys. 
{\bf B560} 33 (1999); \\
A.\,Denner, S.\,Dittmaier, M.\,Roth and D.\,Wackeroth, Nucl. Phys. 
{\bf B587} 67 (2000); \\
A.\,Denner, S.\,Dittmaier, M.\,Roth and D.\,Wackeroth, Phys. Lett. 
{\bf B475} 127 (2000); \\
A.\,Denner, S.\,Dittmaier, M.\,Roth and D.\,Wackeroth, hep-ph/0101257.
\bibitem{tgc}
ALEPH Collaboration, ALEPH 2001-027 CONF 2001-021 (2001);\\
DELPHI Collaboration, DELPHI 2002-016-CONF-555 (2002);\\
L3 Collaboration, L3 Note 2734 (2002);\\
OPAL Collaboration, OPAL Physics Note PN501 (2002).
\bibitem{crpert}
T. Sj\"ostrand and V.A. Khoze, Z. Phys. {\bf C62} 281 (1994).
\bibitem{bec}
ALEPH Collaboration, R. Barate \etal, Phys. Lett. {\bf B478} 50
(2000);\\
DELPHI Collaboration, DELPHI-2002-032-CONF-566 (2002);\\
L3 Collaboration, P. Achard \etal, Phys.\ Lett.\ B {\bf 547} 139
(2002).
\bibitem{cr}
The LEP Collaborations ALEPH, DELPHI, L3 and OPAL and the LEP W
Working Group, LEPEWWG/FSI/2002-01, and references therein.
\bibitem{ski}
T.~Sj\"ostrand and V.A.~Khoze, Z. Phys. {\bf C62} 281 (1994);
Phys. Rev. Lett. {\bf 72} 28 (1994).
\bibitem{arcr}
L.~L\"onnblad, Z. Phys. {\bf C70} 107 (1996);\\
L.~L\"onnblad, Comput. Phys. Comm. {\bf 71} 15 (1992).
\bibitem{mwlep}
The LEP Collaborations ALEPH, DELPHI, L3 and OPAL and the LEP W
Working Group, Note LEPEWWG/MASS/2002-02 (2002) and references
therein.
\bibitem{mwtev}
CDF Collaboration, F.~Abe \etal, Phys.~Rev.~Lett. {\bf 65} 2243 (1990);\\
CDF Collaboration, F.~Abe \etal, Phys.~Rev. {\bf D43} 2070 (1991);\\
CDF Collaboration, F.~Abe \etal, Phys.~Rev.~Lett. {\bf 75} 11 (1995);\\
CDF Collaboration, F.~Abe \etal, Phys.~Rev. {\bf D52} 4784 (1995);\\
A.~Gordon, presented at XXXIInd Rencontres de Moriond, Les Arcs, 
16-22 March 1997; \\
D\O\ Collaboration, S.~Abachi \etal, 
Phys. Rev. Lett. {\bf 84} 222 (2000).
\bibitem{gwtev}
CDF Collaboration, T. Affolder \etal, Phys. Rev. Lett. {\bf 85} 3347
(2000); \\
D\O\ Collaboration, V.M. Abazov \etal, Phys. Rev. {\bf D66} 032008 (2002)
\bibitem{nutev}
NuTeV Collaboration, G.~P.~Zeller {\it et al.},  
Phys.\ Rev.\ Lett.\  {\bf 88} 091802 (2002).
\bibitem{apv}
C.S. Wood \etal., Science {\bf 275} 1759 (1997);\\
S.C. Bennett and C.E. Wieman, Phys. Rev. Lett. {\bf 82} 2484 (1999).
\bibitem{zfitter}
D.~Bardin \etal, Z.~Phys. {\bf C44} 493 (1989);
Comp.~Phys.~Comm. {\bf 59} 303 (1990);
Nucl.~Phys. {\bf B351} 1 (1991);
Phys.~Lett. {\bf B255} 290 (1991) and CERN-TH 6443/92 (1992);  the most
recent version of ZFITTER (6.21) is described in
Comp.~Phys.~Comm. {\bf 133} 229 (2001).
\bibitem{topaz}
G.~Montagna \etal, Comput. Phys. Commun. 
{\bf 117} 278 (1999);\\
{\tt http://www.to.infn.it/$\sim$giampier/topaz0.html~}.
\bibitem{higgs}
ALEPH Collaboration, R.~Barate \etal,
Phys. Lett. {\bf B 526} 191 (2002);\\
DELPHI Collaboration, DELPHI-2002-041-CONF-575 (2002);\\
L3 Collaboration, M. Acciarri \etal, Phys. Lett. {\bf B517} 319
(2001);\\
OPAL Collaboration, G. Abbiendi \etal, CERN-EP-2002-059 (2002).
\bibitem{haleph}
W.D.\,Schlatter for the ALEPH Collaboration, ``ALEPH Status Report'',
presented at the CERN LEPC, 5 September 2002.
\bibitem{lephiggs}
ALEPH, DELPHI, L3 and OPAL Collaborations and the LEP Working Group
for Higgs Boson Searches, LHWG Note/2002-01 (2002).

\end{thebibliography}
\end{document}